\RequirePackage{fix-cm}
\RequirePackage{silence}
\documentclass[pdflatex,sn-mathphys]{sn-jnl}

\jyear{2023}%

\theoremstyle{thmstyleone}%
\usepackage{verbatim}

\usepackage{setspace}
\usepackage{url}
\usepackage[export]{adjustbox}
\usepackage{longtable}

\usepackage[T1]{fontenc}
\usepackage{array}
\usepackage{enumitem}
\usepackage[ruled,vlined,algo2e]{algorithm2e}
\theoremstyle{thmstyleone}%

\usepackage{graphicx}
\usepackage{float}
\SetKwInput{KwInput}{Input}                
\SetKwInput{KwOutput}{Output} 
\usepackage{lineno,hyperref}
\modulolinenumbers[5]
\usepackage{setspace}
\usepackage{amsmath,amsfonts,amssymb}
\usepackage{adjustbox}
\usepackage{listingsutf8}
\usepackage[utf8]{inputenc}

\begin{document}

\title[AMD-FCG]{AMD-FCG: An Enhanced Function Call Graph Dataset with Integrated Topological Features for Malware Detection and Classification}

\author*[1]{\fnm{Parthajit} \sur{Borah}}\email{parthajit.borah@nfsu.ac.in}

\author[1]{\fnm{Sakshi Singh}}
\equalcont{These authors contributed equally to this work.}

\author[2]{\fnm{D.K. Bhattacharyya} }
\author[3]{\fnm{J.K. Kalita} }

\affil*[1]{\orgdiv{School
		of Cyber Security and Digital Forensics}, \orgname{National Forensic Sciences University,Guwahati}}
\affil[2]{\orgdiv{Department
		of Computer Science and Engineering}, \orgname{Tezpur University}, \orgaddress{ \city{Tezpur}, \postcode{784028}, \state{Assam}, \country{India}}}
\affil[3]{\orgdiv{Computer Science, College of Engineering and Applied Science}, \orgname{University of
Colorado,Colorado Springs}, \orgaddress{ \postcode{CO 80933-7150}, \state{Colorado}, \country{USA}}}


\abstract{As malware illustrates a complex structure and behavior, detection of these has been a significant challenge in the domain of cybersecurity along with related services in daily life. So, it becomes crucial to have a reliable and adaptive solution to address the issue. Among the several detection methods developed over the years, one of the most reliable ones is studying and analyzing the structural and behavioral patterns of malware. These patterns of sophisticated malware can be obtained with the help of Function Call Graphs (FCGs). However, to effectively cover numerous groups of families of malware, it is required to have a sufficiently large dataset for the system to operate on. In order to ensure accuracy and robustness of the system, the dataset should comprise samples of different malwares and a benign application for secure execution of the detection process. This paper introduces AMD-FCG, an enhanced Function Call Graph dataset integrated with topological features of malwares. The framework enhances the detection procedure, streamlining the workflow for cybersecurity professionals and also eliminating the need for dynamic analysis and extensive processing. Therefore, it can be used to develop and deploy more efficient and innovative malware detection systems. }




\maketitle
\section{Introduction}
The world is witnessing a digital revolution and a rapid shift in technologies every day. With these advancements--data, services, and systems became essential for the efficient workflow of all domains. However, these factors continuously face significant threats from malware and malware-based attacks. As malware carries ever-evolving sophistication along with complex and advanced methodologies to circumvent security measures, it becomes essential to regard this as an intricate problem and craft robust security mechanisms. The primary agenda behind malware-based attacks is to damage data, compromise information, disrupt services, and in some cases gain unauthorized access to critical systems. Modern cyberthreats encompass a wide range from traditional viruses and worms to advanced and adaptive malware such as ransomware and spyware. \cite{aslan2020comprehensive} \cite{kramer2010general}
 \\
Malware-based attacks are carried out with the intent to conduct unauthorized activities like data theft, operational disruptions, and espionage, leading to significant financial losses and reputational damage. The modern threat actors adopt advanced techniques such as code obfuscation, packing, and encryption to evade traditional detection systems. \cite{you2010malware} \cite{or2019dynamic}. These techniques conceal the true nature, dynamics, and functionality of malware, thereby complicating the identification of malicious software. Therefore, defense mechanisms designed to counter these sophisticated attacks must demonstrate strong resilience while simultaneously possessing the ability to accurately detect even the most intelligently disguised malware. In general practice, advanced security systems actively combine static and dynamic analysis, where the principle behind dynamic is executing the code in a controlled environment, and static analysis involves examining the code, structure, and properties of software typically without execution. \cite{ye2017survey} \cite{moser2007limits} The analysis process reveals intricate patterns and signatures associated with the malware, which play a significant role in its detection. However, static analysis techniques can be compromised by code obfuscation mechanisms; hence, dynamic analysis is essential for thorough observation of malicious software. \cite{or2019dynamic}. Even though such analysis techniques are effective, there are several technical snags. Firstly, dynamic analysis techniques are resource intensive, and for accurate interpretation of malicious behaviors, it requires extensive feature engineering. Secondly and most importantly, such techniques can be bypassed by malware which has the ability to detect virtualized environments. To overcome these challenges, advanced malware data representation techniques are the need of the hour. These novel techniques help in the creation of an efficient and more compact structural representation of malware behavior. Such representations facilitate quicker processing and reduce the need of extensive feature engineering.  Subsequently, malware analysis using such representations is more effective and is less time consuming. This aids in the development of accurate and adaptive defense systems enabling real-time threat detection and response.  Additionally, such systems are also robust in nature against rapidly evolving malware variants, significantly improving overall cybersecurity.\\
Recent upsurge in malware attacks have dramatically given rise to grave threats worldwide. These threats are not confined to one geographical boundary alone and may be effortlessly launched from one faraway region to another. Unsuspecting users of vulnerable and insecure systems end up being victims as malicious software are delivered to the platforms uncannily through various means.

\subsection{Contribution}
The primary goal of this paper is to report the development of an enhanced malware dataset, named AMD-FCG, which integrates both Function Call Graph (FCG) data and topological features for advanced malware detection and classification. AMD-FCG produces a detailed and distinct dataset with the help of various static analysis techniques. These datasets further consist of 30,000 instances, each with 51 distinguishing classes of malware. Here, the 50 classes contribute to the classification of diverse variants of malware, whereas one class is specified for goodware. In addition, the FCG data features the structural relationships along with operative interactions between function calls inside the malicious program. This interactive record inherits information about commands and system/API functions executed to establish the malicious intent, which provides a unique view of malware behavior. The integrated topological characteristics strengthen the analysis procedure by quantifying the structural properties of graphs, which is critical for classifying different malware families. Furthermore, we release AMD-FCG under a CC-BY license, allowing researchers and practitioners to freely share and adapt the dataset to their specific needs. This open approach promotes collaboration and encourages innovative minds for the development of more effective malware detection techniques within the cybersecurity domain. Researchers interested in accessing this dataset for further studies can request it by emailing the corresponding author at \textcolor{blue}{\texttt{parthajit.borah@nfsu.ac.in}}.

\subsection{Motivation}
With the advent of modern technology, the internet and the interconnected systems have essentially become an integral part of every industry. As many services and solutions are conveniently being brought to our doorstep so are the serious security challenges that come with it.  Malware and malware-based attacks pose a grave threat to the users of vulnerable and insecure systems. Sophisticated malware attacks are not only hard to recognize but also are more challenging to detect with traditional security measures. Due to the constant emergence of such attacks, effective and resilient detection mechanisms are the need of the hour. A key factor in designing such mechanisms is the quality of the data used for training and testing purposes. Especially, the data should be able to portray varied and complex real world attack scenarios.  Lack of such quality datasets may lead to unwanted gaps in the security provided by the detection mechanisms. The prime focus of this article is to highlight the importance of this issue by introducing a new, improved dataset designed to enhance the efficacy of malware detection systems, helping to better protect against malware threats.

\subsection{Organisation of the paper}\label{organisation}
The organization of the paper is as follows. In Section \ref{background}, the background and state-of-the-art methods are discussed in length. Section \ref{methodology} on the other hand presents the methodology used to develop AMD-FCG starting from the data collection process, call graph analysis, and topological feature extraction. Section \ref{results} illustrates the results and experimentation using AMD-FCG. Lastly, we wind up with the overall concluding remarks in Section \ref{cf}. 
\section{Background and State of the Art}\label{background}
Malware is a malicious program coded in such a way that it damages data, computer systems, and networks as well. These malicious software are developed by attackers to steal sensitive information and access critical systems without the permission of the system's owner. \cite{aslan2020comprehensive} \cite{kramer2010general}. Malware is emerging as a complex problem and is causing immense damage to various networks as well as the internet. They are in various forms and structures, such as Adware, Trojans, Backdoors, Worms, Bots, Rootkits, Downloaders, Ransomware, and Viruses \cite{chakkaravarthy2019survey} \cite{egele2008survey}. Malware-based attacks are diverse; some attacks are straightforward with a single stage, while others are intricate and multi-staged with various malware as entities to accomplish the malicious goals. In addition, complexity and sophistication update the malicious program to a large extent, strengthening malware to trick detection systems. Simple attacks hold a single piece of malicious code to breach systems and steal information, whereas advance attacks inherit a complex code structure along with intelligently ordered steps to fulfill the hidden agenda. In order to counter these attack vectors, it is important to examine the behavior of malware through a detailed analysis. This can be done by accumulating and uncovering the risk factor and the intended target of the malware. 

 Malware analysis plays a crucial role in recognizing indicators of compromises, as they are further used to train and develop detection models. \cite{yao2019review}. The four core techniques in malware analysis are static, dynamic, automated, and manual code reversing. \cite{or2019dynamic}. Each technique provides strong characteristics with a unique view of malware, leading to a better understanding and reduction in the impact of malware threats. The representation of malware data plays an important role in strengthening defense systems to battle against various malware and malware-based attacks. Apart from just boosting the process of malware detection, it also counters the impact of the unauthorized access. There are various formats to represent the malware data; each format is crafted to deal with specific aspects of threat analysis, and improving the security mechanisms. These formats serve as essential input data for malware defense models. Typically, malware data can be categorized into several distinct types: tabular data, image data, graph data, sequence data, and text data. 

From the mentioned approaches, one of the most reliable ones is examining the structural patterns of malware, and these patterns are captured with the help of function call graphs (FCGs). FCGs give a detailed mapping of all the interactions that occurred between the functions inside the program. This provides a rich collection of information that is crucial in identifying malicious activity. While FCGs are fundamental for revealing these behavioural patterns, they often lack in providing additional contextual information that strengthens the distinguishing ability of the graph. However, this gap can be filled by combining topological features of malware, which gives a broader set of characteristics of the graph, and thus enhancing the overall robustness of malware detection. 

There are a limited number of publicly available function call graph datasets of malware. And from the very few, only one dataset is currently accessible, and that is MalNet \cite{freitas2020large}. These datasets often contain only the function call graphs, where the functions are denoted by numerical identifiers instead of their actual names, which is a bottleneck for data interpretability. In general, they also lack integration of the topological features, which is required to obtain broader structural characteristics of the graph. This bounds the richness of analysis and the scope development of more advanced detection models.

\section{Methods}\label{methodology}
\begin{figure}[t]
	\centering
	\includegraphics[width=0.7\linewidth]{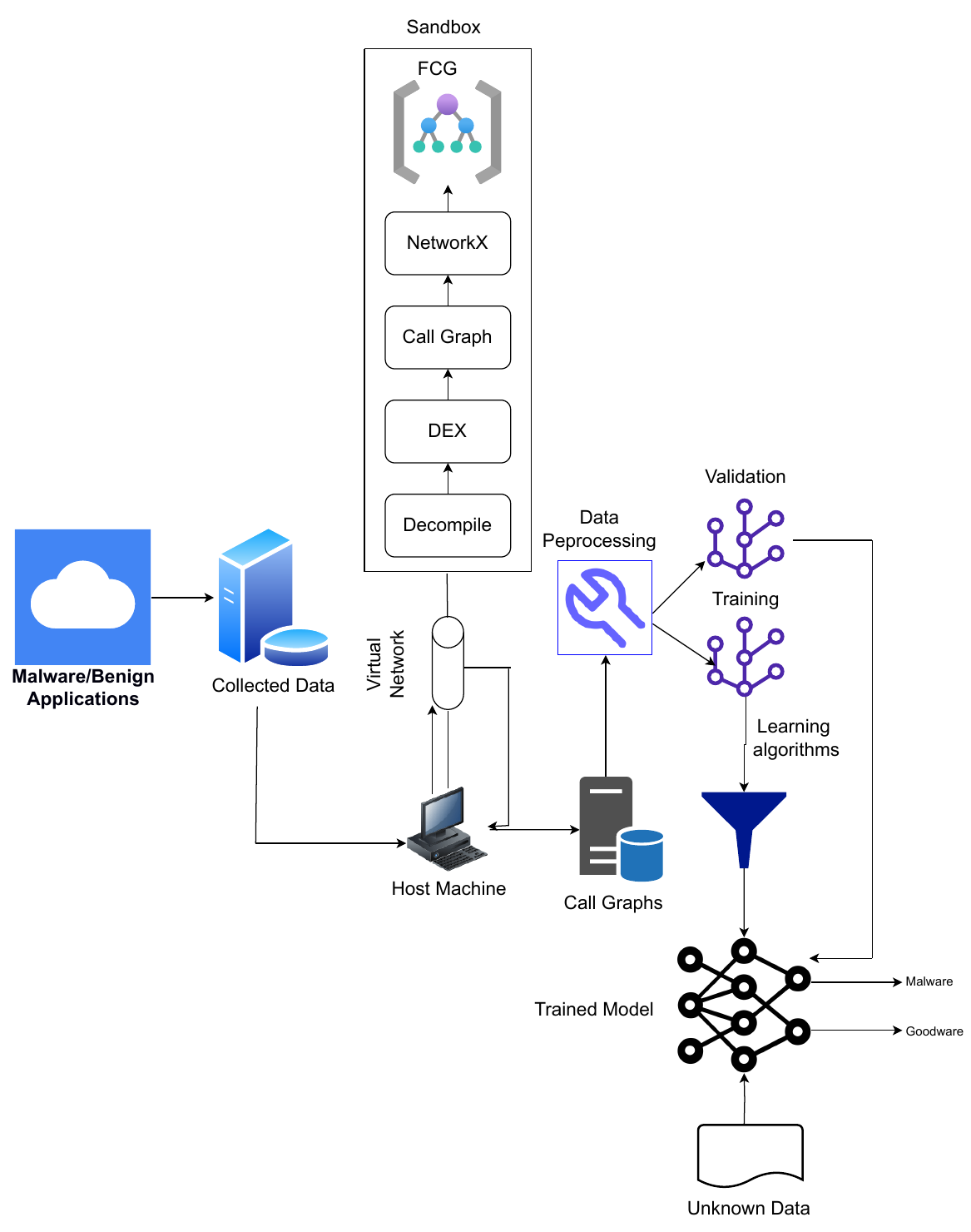}
	\caption{AMD-FCG Generation and Evaluation Framework}
	\label{tundrodfcg}
\end{figure}

 This section covers multiple processes involved in the generation of AMD-FCG's integrated dataset. At first, the Function Call Graphs (FCGs) are produced, and then the topological features are extracted from these graphs. The development of AMD-FCG includes a sequence of comprehensive and rigorous processes, which ensures accurate generation of FCGs and accumulation of meaningful topological features. The proposed framework of AMD-FCG, illustrated in Figure \ref{tundrodfcg}, shows the overall process from dataset creation to validation. The framework contains various processes such as data collection, feature extraction and integration, and model validation.
\subsection{Data Collection}
We collected 10,000 benign Android apps from Google Play Store and AndroZoo as goodware. Moreover, we gathered a total of 20,000 raw Android malware files classifying 50 malware families from \cite{wei2017deep} \cite{borah2020malware} for malware binaries to obtain a comprehensive and extensive data archive. The extracted Android applications are saved in a database server for further processing and analysis. The rich collection of data is used to create Function Call Graphs (FCGs) and gather topological features, which is a crucial step towards better malware detection and classification for the AMD-FCG dataset. 
\subsection{Function Call Graph Extraction}
A series of methodical steps are involved in extracting and analyzing the function calls from Android applications. These steps involve the process of decompilation, analysis, and representation of the application's internal structure in the form of a function call graph. The following outlines the process in detail.

\subsubsection{APK analysis}
The first step is to decompile the Android binaries using the reverse engineering tool Androguard\footnote{https://androguard.readthedocs.io/en/latest/intro/index.html}. With the help of Androguard, the DEX (Dalvik Executable) files, which contain the compiled code of the Android application, are extracted and parsed to extract the Dalvik bytecode. Dalvik bytecode, a register-based instruction set used by the Android Runtime (ART) or Dalvik Virtual Machine, is then analyzed to understand each instruction's purpose and potential impact on the program flow. Additionally, the AndroidManifest.xml file is analyzed to gain insights into the app's structure, permissions, and components.

\subsubsection{Methods Extraction}

Following the analysis of the Dalvik bytecode, the next step involves finding and organizing all the functions or methods used within the application. This includes methods written by the app's developers, as well as those provided by the Android system, which allows the app to interact with different features of the device, such as the user interface, data storage, or hardware features like the camera or GPS. It also includes methods from any third-party libraries the app uses. All these methods give us a clear understanding of how the app works and how it interacts with the Android system and other software libraries.

\subsubsection{Control Flow Analysis}
In this stage, the process of analyzing the bytecode of each method to discover the flow of interaction between the methods is introduced. It is accomplished by various types of method calls like direct, virtual, interface, and super method calls. The relationship between the calling method (caller) and the method being called (callee) is systematically recorded for each method invocation. Besides just tracing these method calls we also analyze the flow of control inside each method with an eye towards understanding conditional branches, loops, and exception handling. This method gives a detailed view of the execution paths in the application, which helps in understanding how the app works or behaves.

\subsubsection{Graph Construction}
The relations between the methods are used to build a directed graph . Each node of the graph is a method and each edge of the graph is a method call or control flow between methods. The graph is stored as an adjacency list for efficient storage and traversal. The numeric identifiers are used instead of the actual method names to protect the sensitive information. This results in a translation mapping between the original method identifiers and the masked ones. This step guarantees that sensitive method names are protected and it still allows comprehensive study of structure and behavior of the application.

\subsubsection{Visualization}
The final step is to visualize the call graph using Graphviz. This visualization gives a graphical representation of the abstract data and provides a whole view of the application structure. It reduces complex relationships to an easy to understand visual map. It gives a better view of the interaction and collaboration of different parts of the application.

\begin{algorithm2e}[H]
    \SetAlgoLined
    \KwInput{list\_of\_apks}
    \KwOutput{list\_of\_call\_graphs}
    
    \ForAll{apk in list\_of\_apks} {
        a, d, dx $\gets$ AnalyzeAPK(apk)\;
        call\_graph $\gets$ dx.get\_call\_graph()\;
        node\_to\_number $\gets$ Encode each function name in call\_graph to a unique number\;
        numbered\_graph $\gets$ label call\_graph using node\_to\_number\;
        return (numbered\_graph, call\_graph);
    }
    
    \caption{Process APKs and Generate Call Graphs}
\end{algorithm2e}

Table \ref{snapfcg} displays the snapshot of the Function Call Graph of Airpush malware, including the source and target functions along with descriptions of the calls.

\begin{table}[]
\caption{Snapshot of Function Call Graph (FCG) for Airpush Malware}
\label{snapfcg}
\scalebox{0.7}{
\begin{tabular}{|l|l|l|}
\hline
Source Function & Target Function & Description \\ \hline
\begin{tabular}[c]{@{}l@{}}HttpPostDataTask.\textless{}init\textgreater\\ (Context, List, String, Listener)\end{tabular} &
  Util.printDebugLog(String) &
  \begin{tabular}[c]{@{}l@{}}Constructor of HttpPostDataTask calls the static method \\ printDebugLog in the Util class.\end{tabular} \\ \hline
\begin{tabular}[c]{@{}l@{}}HttpPostDataTask.\textless{}init\textgreater\\ (Context, List, String, Listener)\end{tabular} &
  AsyncTask.\textless{}init\textgreater{}() &
  \begin{tabular}[c]{@{}l@{}}Constructor of HttpPostDataTask calls the constructor \\ of AsyncTask from Android OS.\end{tabular} \\ \hline
\begin{tabular}[c]{@{}l@{}}HttpPostDataTask.\textless{}init\textgreater\\ (Context, List, String, Listener)\end{tabular} &
  StringBuilder.toString() &
  \begin{tabular}[c]{@{}l@{}}HttpPostDataTask constructor includes a call to \\ StringBuilder's toString method.\end{tabular} \\ \hline
Airpush\$3.onTaskComplete(String) &
  \begin{tabular}[c]{@{}l@{}}AsyncTaskCompleteListener\\ .lauchNewHttpTask()\end{tabular} &
  \begin{tabular}[c]{@{}l@{}}onTaskComplete method in Airpush\$3 calls\\ launchNewHttpTask  in AsyncTaskCompleteListener.\end{tabular} \\ \hline
\end{tabular}}
\end{table}

\subsection{Importance of Topological Features in Malware Detection}

   Function Call Graphs (FCGs) topological features are robust tools to identify malware since they encode the structural and behavioral features of software even when malware authors attempt to hide their intentions using obfuscation techniques \cite{zhang2018dalvik} \cite{polymorphic}. Some features such as Betti numbers, homology groups and the Euler characteristic can be used to identify important properties of the graph such as how functions are connected, if there are loops and how complex the graph is \cite{wu2018software}. These features are especially useful in malware detection as they are not influenced by tricks such as renaming functions, inserting useless code or changing control flow. Betti numbers can uncover patterns such as loops, which are frequently used in encryption routines or repetitive malicious behaviors (see \cite{pranav2017topology}).

\subsection{Topological Feature Extraction}

Topological analysis involves a systematic approach to uncovering and understanding the shape and structure of data. Various topological features are extracted, providing insights into the data's underlying patterns and properties. Below, we explore these topological features in detail.

\begin{enumerate}
    \item {Simplices}

A \textbf{simplex} is the simplest type of geometric object that can exist in any given dimension. It generalizes the concept of points, line segments, and triangles to higher dimensions \cite{jonsson2008simplicial} \cite{munkres2018elements}.

\begin{itemize}
    \item \textbf{0-simplex:} A point.
    \item \textbf{1-simplex:} A line segment, formed by connecting two points.
    \item \textbf{2-simplex:} A triangle, formed by connecting three non-collinear points.
\end{itemize}

Mathematically, a $k$-simplex is defined as:

\begin{equation}
\sigma = \left\{ x \in \mathbb{R}^d : x = \sum_{i=0}^{k} t_i p_i, \right.
\end{equation}
\[
\left. \text{where } \sum_{i=0}^{k} t_i = 1, \text{ and } t_i \geq 0 \text{ for all } i \right\}
\]

where $p_0, p_1, \ldots, p_k$ are the vertices of the simplex, and $t_i$ are the barycentric coordinates.

A \textbf{simplicial complex} represents all the connections (edges) between nodes in a call graph. 

\begin{itemize}
    \item \textbf{Nodes} (functions, methods, or procedures in the call graph) are represented as \textbf{0-simplices}.
    \item \textbf{Edges} (calls between these nodes) are represented as \textbf{1-simplices}.
    \item If three or more nodes are fully interconnected (i.e., each node is connected to every other node), this relationship can be represented by a \textbf{2-simplex} (a triangle) or higher-dimensional simplices for more complex connections.
\end{itemize}

\begin{figure}[tbp]
	\centering
	\includegraphics[width=0.8\linewidth]{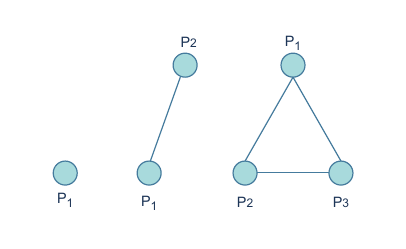}
	\caption{Examples of simplices of dimensions 0, 1, and 2 }
	\label{tundrodfcg1}
\end{figure}

\item {Betti Numbers }

Betti numbers \( \beta_k \) are topological invariants used to describe the \( k \)-dimensional features of a topological space. Specifically, \( \beta_0 \) counts the number of connected components, \( \beta_1 \) counts the number of independent loops, and \( \beta_2 \) counts the number of cavities or voids \cite{carlsson2009topology} \cite{munkres2018elements}.

Mathematically, the \( k \)-th Betti number is defined as:
\begin{equation}
    \beta_k = \text{dim}(Z_k) - \text{dim}(B_k)
\end{equation}

where \( Z_k \) is the space of \( k \)-cycles and \( B_k \) is the space of \( k \)-boundaries.

For example in Figure \ref{tundrodfcg1}:

\begin{itemize}
    \item $\beta_0$ (Connected Components): $\beta_0 = 1$
    
    The simplicial complex has only one connected component, as all vertices ($P_1$, $P_2$, and $P_3$) are part of the same structure.

    \item $\beta_1$ (Loops or Cycles): $\beta_1 = 1$
    
    There is one loop (the triangle formed by \( p_1 \), \( p_2 \), and \( p_3 \)) present in the complex. 

    \item $\beta_2$ (Voids): $\beta_2 = 0$
    The complex does not contain any 2-dimensional voids 
\end{itemize}

\item {Euler characteristic}

The Euler characteristic is a fundamental topological invariant that provides a single number summarizing the shape or structure of a simplicial complex \cite{stillwell2012classical}. For a call graph, it is calculated as:

\begin{equation}
    \chi = V - E + F
\end{equation}

where \( V \) is the number of vertices (0-simplices), \( E \) is the number of edges (1-simplices), and \( F \) is the number of filled triangles (2-simplices). This invariant gives an overview of the graph's topological complexity.

\item {Persistence Features}

Persistence features are key metrics in topological data analysis that provide insight into the structure and distribution of topological features within a persistence diagram \cite{rucco2016characterisation}.
\begin{itemize}
    \item {Persistence Entropy}

Persistence entropy \( E(D) \) measures the unpredictability or complexity of the features' lifetimes in a persistence diagram. It is defined as:
\begin{equation}
E(D) = - \sum_{i \in I} p_i \log(p_i)
\end{equation}
where
\[
p_i = \frac{d_i - b_i}{L_D}, \quad \text{and} \quad L_D = \sum_{i \in I} (d_i - b_i)
\]
with \( b_i \) and \( d_i \) representing the birth and death times of features, respectively.

\item {Mean Persistence}

Mean persistence \( \mu(D) \) represents the average lifespan of the features in the persistence diagram, calculated as:
\begin{equation}
\mu(D) = \frac{1}{|I|} \sum_{i \in I} (d_i - b_i)
\end{equation}
where \( |I| \) is the total number of features, and \( b_i \) and \( d_i \) are their birth and death times.
\end{itemize}

\item{Morse Analysis}

Morse analysis is a mathematical method used to study the topology of manifolds by examining the critical points of a smooth function on the manifold. For a graph \( G \), Morse analysis identifies and categorizes critical points based on their local neighborhoods \cite{matsumoto2002introduction} \cite{milnor1963morse}.

\begin{equation}
\text{Critical Point Category} = 
\begin{cases} 
\text{maximum} & \text{if } \deg(v) > \deg(u) \text{ for all } u \in N(v) \\
\text{minimum} & \text{if } \deg(v) < \deg(u) \text{ for all } u \in N(v) \\
\text{saddle} & \text{if } \deg(v) \text{ has both neighbors with} \\
\text{} & \text{higher and lower degree} \\
\text{isolated} & \text{if } \deg(v) = 0 
\end{cases}
\end{equation}

Here, \( \deg(v) \) represents the degree of vertex \( v \) and \( N(v) \) is the set of neighbors of \( v \). By analyzing the critical points, the topological features of the graph are extracted, which will help in understanding its structure and characteristics.

\item{Homology and Cohomology}

Homology is a concept in algebraic topology that studies the topological features of a space, such as connected components, holes, and voids \cite{zomorodian2004computing} \cite{munkres2018elements}. For a simplicial complex \( K \), the \textbf{n-th homology group} \( H_n(K) \) is defined as:
\begin{equation}
H_n(K) = \frac{Z_n(K)}{B_n(K)}
\end{equation}
where \( Z_n(K) \) is the group of n-cycles (chains with no boundary) and \( B_n(K) \) is the group of n-boundaries (boundaries of (n+1)-chains).

Cohomology is a dual concept to homology and provides an algebraic structure that encodes the information of the space. For a simplicial complex \( K \), the \textbf{n-th cohomology group} \( H^n(K) \) is given by:
\[
H^n(K) = \text{Hom}(H_n(K), \mathbb{Z})
\]
This group consists of homomorphisms from the n-th homology group to the integers \( \mathbb{Z} \), capturing the dual structure of the homology.

\item{Reeb Graph}

The Reeb graph size feature is defined as the number of nodes in the largest connected component of the Reeb graph, which captures the evolution of level sets of a continuous function on a graph. It provides a simplified representation of the topology of the graph \cite{carriere2018statistical}.

\begin{equation}
\text{Reeb\_Graph\_Size}(G) = \max\bigl(|\text{Connected\_Components}(G, f)|\bigr)
\end{equation}

where \( G \) is the graph and \( f \) is a continuous function defined on \( G \).

\end{enumerate}

\begin{algorithm2e}[H]
\caption{Extract Features from FCGs}
\KwIn{List of FCGs $\mathcal{G}$}
\KwOut{Feature vector for each FCG}
\ForEach{$g \in \mathcal{G}$}{
    $e \leftarrow$ read\_graph($g$)\;
    $v_0 \leftarrow$ count\_nodes($e$)\;
    $v_1 \leftarrow$ count\_edges($e$)\;
    $v_2 \leftarrow$ count\_triangles($e$)\;
    $b_0, b_1 \leftarrow$ calculate\_betti\_numbers($e$)\;
    $ec \leftarrow$ calculate\_euler\_characteristic($v_0, v_1, v_2$)\;
    $st \leftarrow$ generate\_vietoris\_rips($e$)\;
    $pe, mp \leftarrow$ calculate\_persistence($st$)\;
    $hg \leftarrow$ calculate\_homology\_groups($st$)\;
    $coh \leftarrow$ calculate\_co\_homology($st$)\;
    $mt \leftarrow$ calculate\_morse\_features($e$)\;
    $rg \leftarrow$ calculate\_reeb\_graph($e$)\;
    \Return $\{v_0, v_1, v_2, b_0, b_1, ec, pe, mp, hg, coh, mt, rg\}$
}
\end{algorithm2e}

\section{Experiments and Results}\label{results}
All experiments has been implemented in Python using a Dell Precision 7810 workstation with 2x Intel Xeon (R) W-2145 comprising 8 cores, 64GB RAM, NVIDIA RTX 4090 GPU with 24GB VRAM, and Ubuntu OS. Materials used, preprocessing carried out, and performance achieved are discussed next.

\subsection{Dataset Characteristics}
\begin{figure}[htbp]
	\centering
\includegraphics[width=0.3\linewidth]{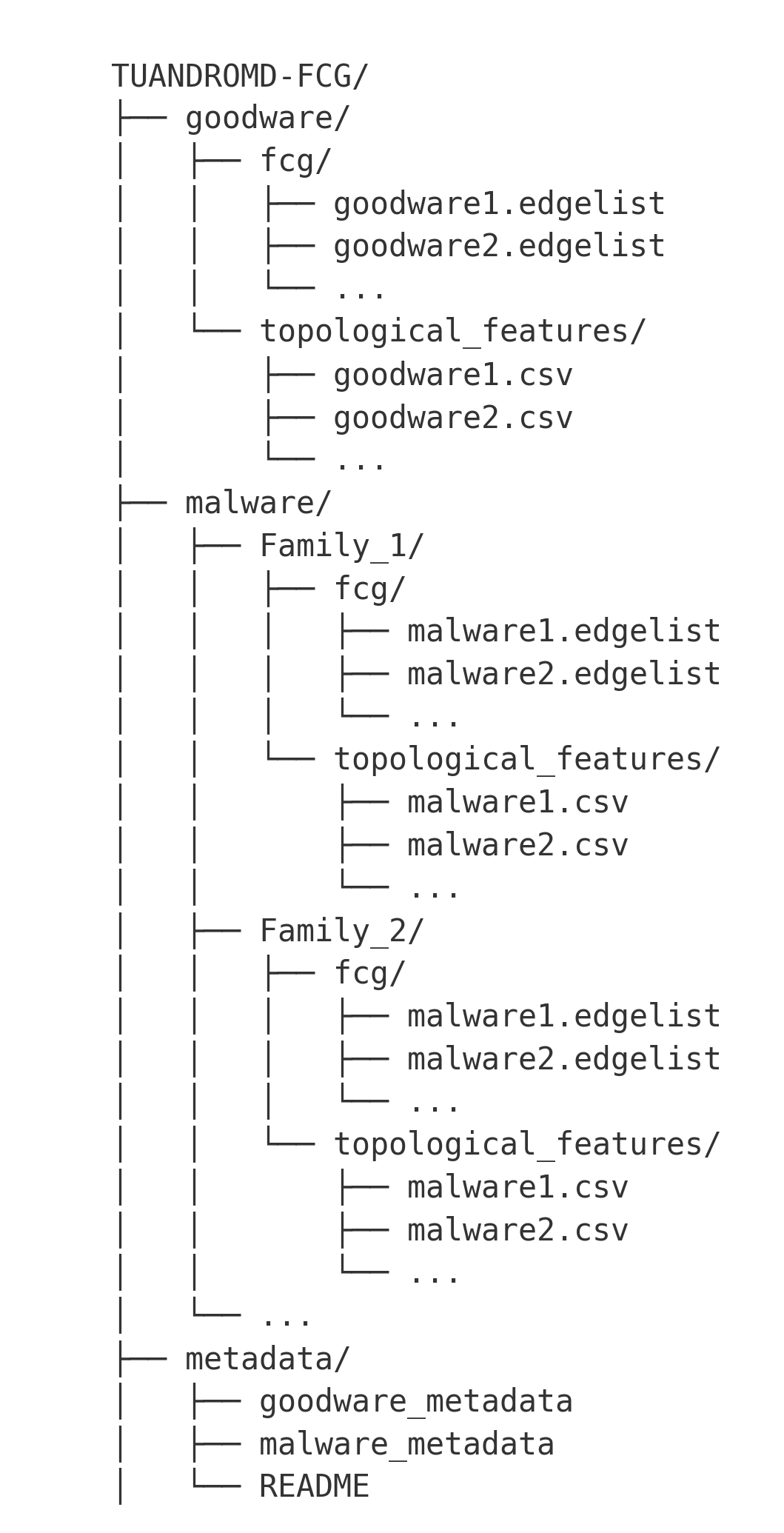}
	\caption{Directory structure of AMD-FCG}
	\label{DIRECT}
\end{figure}

\begin{figure}[htbp]
	\centering
	\includegraphics[width=0.8\linewidth]{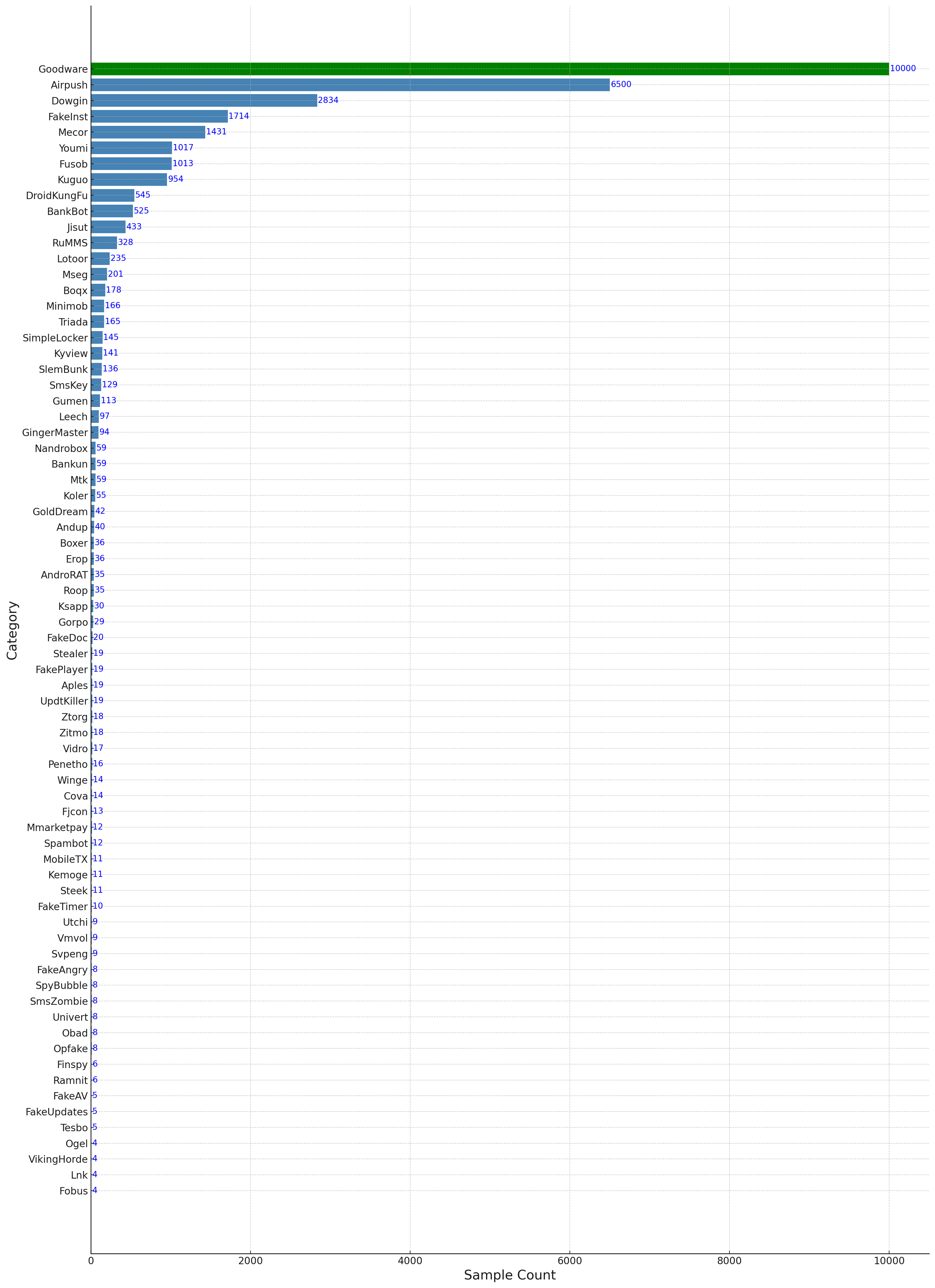}
	\caption{Class Distribution of AMD-FCG}
	\label{classd}
\end{figure}

\begin{figure}[htbp]
	\centering
	\includegraphics[width=0.8\linewidth]{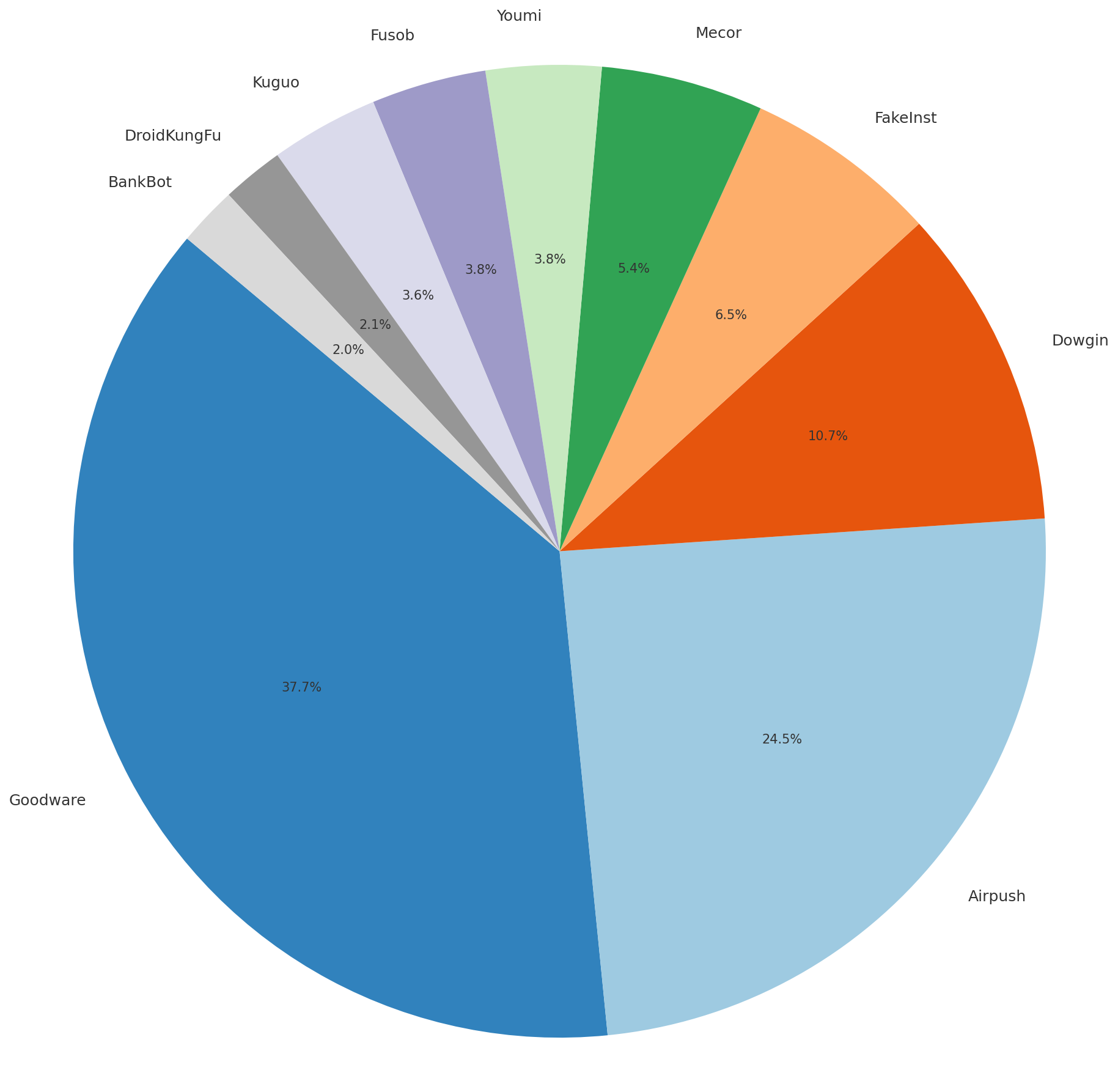}
	\caption{Top 10 categories of AMD-FCG}
	\label{top10}
\end{figure}

The AMD-FCG  dataset consists of 30,000 instances of malware and goodware samples. It comprises 20,000 malware instances, and 10,000 goodware instances. The dataset is categorised into 51 distinct classes, with 50 classes dedicated to various types of malware and one class for goodware. The detailed dataset statistics are presented in Table~\ref{tab:dataset_statistics}. To provide a visual representation of the data distribution, Figure~\ref{classd} illustrates the overall class distribution, while Figure~\ref{top10} shows the top 10 categories within the dataset. 
The figure \ref{DIRECT} illustrates the directory structure, showcasing how the AMD-FCG dataset is organized.

\begin{table}[htbp]
\centering
\caption{Dataset Statistics}
\label{tab:dataset_statistics}
\begin{tabular}{lr}
\hline
\textbf{Characteristic} & \textbf{Count} \\
\hline
Total Instances & 30,000 \\
Malware Instances & 20,000 \\
Goodware Instances & 10,000 \\
\hline
Total Classes & 51 \\
Malware Classes & 50 \\
Goodware Class & 1 \\
\hline
\end{tabular}
\end{table}

\subsection{Performance Evaluation}
For performance evaluation, we considered five Graph Neural Network (GNN) models—GCN, GraphSAGE, GIN, GAT, and GraphConv—and observed their performance on the Function Call Graph (FCG) dataset. The results of these models on the FCG dataset are reported in Table \ref{tab:gnn_performance}. On the FCG dataset, GIN achieved the highest accuracy of 77.1\%, while GraphSAGE and GCN showed slightly lower performances with accuracies of 75.2\% and 73.5\%, respectively. GAT and GraphConv attained accuracies of 70.8\% and 72.0\%, respectively, on the same dataset.
\begin{table}[htbp]
\centering
\caption{Performance of GNN Models on FCG Dataset}
\label{tab:gnn_performance}

\begin{adjustbox}{max width=\linewidth}
\begin{tabular}{lcccc}
\hline
\textbf{Model} & \textbf{Accuracy (\%)} & \textbf{Precision(\%)} & \textbf{Recall(\%)} & \textbf{F1-Score (\%)} \\
\hline
GCN & 73.5 & 72.8 & 74.0 & 73.4 \\
GraphSAGE & 75.2 & 74.5 & 75.7 & 75.1 \\
GIN & 77.1 & 76.4 & 77.5 & 76.9 \\
GAT & 70.8 & 70.2 & 71.1 & 70.6 \\
GraphConv & 72.0 & 71.3 & 72.5 & 71.9 \\
\hline
\end{tabular}
\end{adjustbox}
\end{table}

We used traditional machine learning models to evaluate the potential of our dataset for malware classification based on topological features. We chose these models because they have been proven to be effective in handling complex data and capturing the important patterns that differentiate various types of malware. To make sure the assessment is exhaustive and impartial, we have adopted a strategic data partitioning scheme. The dataset was divided into three separate subsets, 70\% was used to train the models, 10\% was used for validation to fine-tune the models and the remaining 20\% was used to test the final performance. We use these classical machine learning methods on our well-partitioned data set of topological features and try to set a benchmark for malware classification performance.

For performance evaluation, we considered five traditional machine learning models—Random Forest, Support Vector Machine, Gradient Boosting, K-Nearest Neighbors, and Logistic Regression on the topological feature dataset.  The results of these models on the topological dataset are reported in Table \ref{tab:ml_performance}. In contrast, when evaluating the topological feature dataset, the traditional machine learning models generally achieved higher accuracy metrics. Gradient Boosting led with an accuracy of 85.0\%, followed by Support Vector Machine and Random Forest, with accuracies of 84.1\% and 83.2\%, respectively. K-Nearest Neighbors and Logistic Regression showed competitive results, with accuracies of 79.8\% and 81.0\%, respectively. These results indicate that while GNN models like GIN perform well on graph-structured data such as FCG, traditional machine learning models generally excel on structured datasets with topological features. Gradient Boosting had the highest overall performance on the topological feature dataset, while GIN performed best on the FCG dataset.

\begin{table}[htbp]
\centering
\caption{Performance of Machine Learning Models on Topological Feature Dataset}
\label{tab:ml_performance}

\begin{adjustbox}{max width=\linewidth}
\begin{tabular}{lcccc}
\hline
\textbf{Model} & \textbf{Accuracy (\%)} & \textbf{Precision(\%)} & \textbf{Recall(\%)} & \textbf{F1-Score (\%)} \\
\hline
Random Forest & 83.2 & 82.9 & 83.4 & 83.1 \\
Support Vector Machine & 84.1 & 83.8 & 84.3 & 84.0 \\
Gradient Boosting & 85.0 & 84.7 & 85.2 & 84.9 \\
K-Nearest Neighbors & 79.8 & 79.5 & 80.0 & 79.7 \\
Logistic Regression & 81.0 & 80.7 & 81.2 & 80.9 \\
\hline
\end{tabular}
\end{adjustbox}
\end{table}

\subsection{Limitations and Challenges}
One of the main limitations of the datasets used for malware classification, the Function Call Graph (FCG) dataset and the topological feature dataset, is the problem of class imbalance. This is especially true for the underrepresentation of certain malware classes, which poses difficulties for model training and evaluation. These classes are hard to get enough samples of as they are rare classes and are not easily available in public databases. Augmentation techniques can be used to balance classes in traditional image datasets. However, augmentation of graph-based or topological feature data is not straightforward. Augmentation to artificially increase the number of samples may introduce artifacts or distortions that do not accurately reflect the true structural or behavioral characteristics of the original malware.

For FCG dataset, it is important to keep the original distribution even if it is imbalanced, in order to keep the authenticity and integrity of the malware’s functional behavior. Any augmentation on graph structures may result in inaccurate representations of the malware's execution flow, which could mislead the GNN models during training and evaluation. Similarly, for the topological feature dataset, adding synthetic samples can interfere with the topological patterns needed for accurate classification. Instead, we opted not to employ any augmentation techniques and concentrated on developing models that can robustly deal with imbalanced data.

Furthermore, the collection of samples from underrepresented malware classes is a major challenge. To collect these samples, it is often necessary to deploy advanced honeynets or other specialized techniques, which can be both expensive and resource intensive. This issue can be resolved through collaboration within the cybersecurity community to facilitate safer sharing of malware samples. The development of more sophisticated and cost-effective honeypot technologies can also help to capture a wider range of malware samples to enhance the diversity of the dataset.

Another specific challenge of the topological feature dataset is the requirement of further exploration of diverse feature extraction techniques. The current features work. However, other topological measures can be explored to discriminate subtle variations in malware behaviors that are not captured by the current feature set. This may improve the model's ability to distinguish goodware from similar malware instances, and thus improve classification performance.

\section{Conclusion}\label{cf}
In this work, we present AMD-FCG, a malware dataset enriched with Function Call Graph (FCG) data and topological features for advanced malware detection and classification. The importance of AMD-FCG is that it can reflect the features and the relationships of malware function calls by static analysis. So, it is a valuable resource for researchers. In comparison with the datasets based on the dynamic analysis, AMD-FCG offers a quicker and more efficient way to produce significant data representations for rapid classification and analysis. The dataset is tested by both Graph Neural Network (GNN) and traditional machine learning models. The success of the dataset demonstrates its versatility and effectiveness in the training and benchmarking of the algorithms for malware detection. By making AMD-FCG publicly available, we aim to contribute to the advancement of cybersecurity research and support the development of more robust malware detection systems.


\section*{Declarations}
\subsection*{Conflict of interest}
Conflict of Interest:  On behalf of all authors, the corresponding author states that there is no conflict of interest.
\subsection*{Authors' contributions}

\subsection*{Ethics approval and consent to participate}
Not applicable
\subsection*{Availability of data and material}
Made available on request
\subsection*{Consent for publication}
On behalf of all the authors, “I, the Corresponding Author, declare that this manuscript is original, has not been published before, and is not currently being considered for publication elsewhere.
\subsection*{Funding}
Not applicable



\bibliography{sample-base}


\end{document}